\documentclass[twocolumn,pra,showpacs,superscriptaddress]{revtex4}
\usepackage[pdftex]{graphicx}
\usepackage{amsmath}
\usepackage{amssymb}
\usepackage{subfigure}

\newcommand{\figwidth}{0.98\columnwidth}

\newcommand{\pd}{\phantom{\dagger}}

\renewcommand{\Im}{\operatorname{Im}}

\begin{document}

\title{Lattice modulation spectroscopy of strongly interacting bosons\\
 in disordered and quasi-periodic optical lattices}

\author{G.~Orso}
\affiliation{Laboratoire Physique Th\'{e}orique et Mod\`{e}les Statistiques, Universit\'{e} Paris Sud, Bat. 100, 91405 Orsay Cedex, France}
\author{A.~Iucci}
\affiliation{Instituto de F\'{\i}sica la Plata (IFLP) - CONICET and Departamento de F\'{\i}sica,
Universidad Nacional de La Plata, CC 67, 1900 La Plata, Argentina}
\affiliation{DPMC-MaNEP, University of Geneva, 24 Quai Ernest Ansermet CH-1211 Geneva 4, Switzerland}
\author{M.~A.~Cazalilla}
\affiliation{Centro de F\'{\i}sica de Materiales (CSIC-UPV/EHU). Edificio Korta, Avenida de Tolosa, 72. 20018 San Sebasti\'an, Spain}
\affiliation{Donostia International Physics Center (DIPC), Manuel de Lardiz\'abal 4, 20018 San Sebasti\'an, Spain}
\author{T.~Giamarchi}
\affiliation{DPMC-MaNEP, University of Geneva, 24 Quai Ernest Ansermet CH-1211 Geneva 4, Switzerland}

\begin{abstract}
We compute the absorption spectrum of strongly repulsive one-dimensional bosons in a disordered or quasi-periodic optical lattice. At commensurate filling, the particle-hole resonances of the Mott insulator are broadened as the disorder strength is increased. In the non-commensurate case, mapping the problem to the Anderson model allows us to study the Bose-glass phase. Surprisingly we find that a perturbative treatment in both cases, weak and strong disorder, gives a good description at all frequencies. In particular we find that the infrared absorption rate in the thermodynamic limit is quadratic in frequency. This result is unexpected, since for other quantities like the conductivity in one dimensional systems, perturbation theory is only applicable at high frequencies. We discuss applications to recent experiments on optical lattice systems, and in particular the effect of the harmonic trap.
\end{abstract}
\maketitle
\section{Introduction}

In recent years, developments in the field of ultracold atomic gases have considerably enlarged the possibilities for exploring the physics of strongly correlated systems~\cite{bloch_cold_atoms_optical_lattices_review}. For instance, the study of quantum phase transitions, a subject of continuous theoretical interest, has been strongly stimulated by the experimental observation of the superfluid to Mott insulator transition using optical lattices~\cite{greiner_transition_superfluid_mott}. Indeed, ultracold atom systems offer us an unprecedented
control over the system parameters, which can allow us to ultimately understand the physics of
very hard problems such as the phase diagram of the Hubbard model
in two-dimensions~\cite{ho_attractive_hubbard}.

A particularly interesting and fertile arena is the study of disordered ultra-cold atoms. Random potentials can be introduced in a controlled way by laser beams generating speckle patterns~\cite{lye_speckle_disorder,%
clement_speckle_transport,fort_speckle_expansion,schulte_anderson_localization,chen_phase_coherence,white_disordered_bosons_optical_lattice}, or by loading in an optical lattice  a mixture of two kinds of atoms, one heavy and one light. When the heavy atoms
become randomly localized in the lattice, they will act as  impurities for the lighter atoms~\cite{gavish_disorder_impurities}. Another available technique is to superimpose two optical lattices with incommensurate periodicities, thus generating a quasi-periodic potential~\cite{fallani_bichromatic_shaking}. The quasi-periodic lattices as well as the speckle patterns have been recently used in the experimental efforts to observe the effects of  Anderson localization in dilute Bose gases expanding in highly elongated traps~\cite{aspect_anderson_localization_BEC,roati_anderson_localization_BEC}.
Since Anderson localization is a single-particle effect, the next logical step is to study the interplay of disorder and interactions in strongly interacting ultra-cold atomic systems~\cite{orso_BEC-BCS_crossover,roux_quasiperiodic_phase_diagram,deng_phase_diagram_bichromatic,roscilde_one_dimensional_superlattices}.
The latter may be accessible by tuning interparticle interactions using Feshbach resonances~\cite{roati_feshbach_resonances}, or by loading the atoms in sufficiently deep optical lattices\cite{white_disordered_bosons_optical_lattice} and/or
strongly confining them to low-dimensions in tight traps~\cite{bloch_cold_atoms_optical_lattices_review}.

In the context of the efforts described above, one of the experimental challenges is to observe in ultra-cold gases clear signatures of the theoretically-predicted transition from a superfluid to the Bose-glass phase~\cite{giamarchi_anderson_localization_1D,fisher_bose_glass,delande_compression}. A pioneering step in this direction was recently taken by  the Florence group by using lattice modulation spectroscopy~\cite{fallani_bichromatic_shaking}. This technique consists in heating an ultra-cold gas loaded in a optical lattice by periodically modulating the depth of the lattice~\cite{stoeferle_shaking_fast_tunnability,iucci_shake_bosons_theory,kollath_bosons_shaking_dmrg}. When perturbed in this way,  the gas is driven out of equilibrium and absorbs energy. When the perturbation is switched off,  and after re-thermalization, the broadening of the momentum distribution around zero momentum is taken as a measure of the energy absorbed by the system during the lattice modulation~\cite{stoeferle_shaking_fast_tunnability}. On the theory side, for non-disordered lattices, the calculation of the energy absorption rate due to the lattice modulation was first performed analytically within linear response theory by some of the present authors~\cite{iucci_shake_bosons_theory}.
These results were confirmed and extended beyond linear response using time-dependent density-matrix renormalization group methods, both for the case of bosons~\cite{kollath_bosons_shaking_dmrg} and fermions~\cite{kollath_shake_fermions_DMRG}. More recently, for disordered optical lattices,  the energy absorption rate has been numerically calculated using full diagonalization in small  systems~\cite{hild_shaking_bichromatic,hild_quasimomentum_absorption_superlattice}. Other methods
that have also been discussed in the literature for detecting signatures of the effect of disorder (or quasi-periodicity)
on interacting boson systems in one  dimension focus on the momentum  distribution~\cite{egger_disorder,deng_phase_diagram_bichromatic}
in disordered systems as well as quasi-periodic systems or the expansion dynamics in quasi-periodic
systems~\cite{roux_quasiperiodic_phase_diagram}.

%

Let us consider a one dimensional disordered Bose gas described by the following Hamiltonian:
\begin{align}\label{Hbh}
 H &=-J\sum_{j}(b_{j+1}^{\dag }b_{j}^{\pd}+\text{h.c.}) +\frac{U}{2}\sum_{j}n_{j}(n_{j}-1) \nonumber\\
 & +\sum_{j}( \epsilon _{j}
+ V^\textrm{ho}_j ) n_{j},
\end{align}
where $b_j$ denotes the boson annihilation operator at sites $j$, $n_{j}=b_{j}^{\dag }b_{j}^{\pd}$ being the local density.
Here $J$ and $U$ are the usual parameters of the Bose-Hubbard model  corresponding to the tunneling rate and the onsite repulsion $(U>0)$. The last term in the rhs of Eq.~(\ref{Hbh}) accounts for the presence of both the harmonic trap, $V^\textrm{ho}_j$, and the disorder potential $\epsilon_j$.

The distribution of the on-site energies $\epsilon_j$  in Eq.~(\ref{Hbh}) depends on the specific choice of the random (or pseudo-random) potential. In this work we extensively compare two cases:
\begin{eqnarray}\label{sources}
&a)& \;\epsilon_j \;\textrm{uniformly distributed in}\;\;  [-\Delta ,\Delta ] \\
&b)&  \;\epsilon_j = \Delta \cos(2\pi j \sigma),\;\;\;\; \textrm{$\sigma$  irrational},
\end{eqnarray}
where $\Delta$ measures the strength of the disorder.
From the experimental point of view, speckle patterns and lattice containing heavy atom
impurities can be modeled by case a) whereas
quasi-periodic potentials  obtained by superimposing two
optical potentials with incommensurate periodicities $d_1/d_2=\sigma$ are described
by case b).

In this work we assume that the hopping amplitude $J$ in Eq.~(\ref{Hbh}) is modulated periodically
in time according to  $J(t)=J+\delta J\cos \omega t$, where $\omega$ is the modulation frequency.
We calculate the energy absorption rate  within linear response theory, which is valid for weak lattice modulations $\delta J \ll J$.

We shall restrict ourselves to the strongly repulsive regime  of Eq.~(\ref{Hbh}), where the on-site repulsion is large compared to both the hopping amplitude and the disorder strength  ($U \gg J,\Delta$).
We first discuss systems in the thermodynamic limit by setting $V^\textrm{ho}_j =0$. Effects of the harmonic trapping
potential will be discussed later, in Sect.\ref{sec:sec5}. We also set $\hbar=1$ to simplify the notation.

In the thermodynamic limit both the number $N$ of bosons and the length $M$ of the chain diverge. The ratio $\nu=N/M$ is instead finite and corresponds to the \emph{filling} factor, i.e. the average number of  bosons per lattice site.
We must  distinguish between two physically different situations.
For \textsl{incommensurate fillings} $\nu<1$, the system has gapless excitations. Thus, a good approximation to
the absorption rate at frequencies $\omega \ll U$ can be obtained  by formally taking $U\to +\infty$ and
mapping the resulting hard-core bosons to
non-interacting fermions using a Jordan-Wigner transformation (see \emph{e.g.}
Ref.~\cite{Cazalilla_tonks_gases}).  The  resulting single particle problem can be easily solved
for a given choice of $\epsilon_i$ and $V^\textrm{ho}$ and the energy absorption obtained.
On the other hand,  for \textsl{unit filling} ($\nu=1$) the ground state is a Mott insulator with a gap. In this case,
the elementary excitations are particles and holes~\cite{iucci_shake_bosons_theory}
and in a homogeneous system the absorption can only occur at frequencies $\omega \sim U$.
In the absence of disorder, the response of the system to the lattice modulation has been computed using degenerate perturbation theory within the subspace of particle and hole excitations~\cite{iucci_shake_bosons_theory}.
The same methods can be generalized to deal with disordered case, as shown below.

\begin{figure}[tb]
\begin{center}
\includegraphics[width=\figwidth]{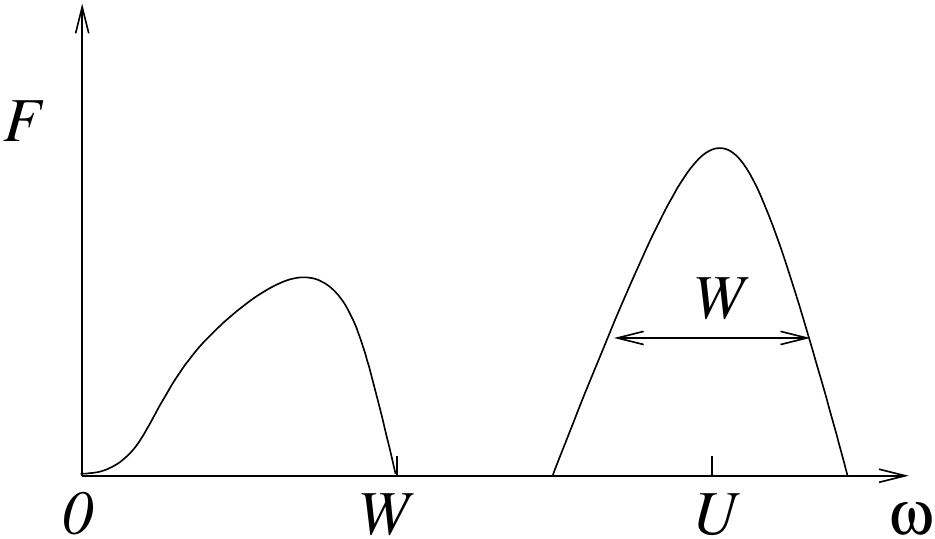}
\end{center}
\caption{
Sketch of the full absorption spectrum at incommensurate filling. Here $W \sim \textrm{max}(4J,2\Delta)$  is the effective bandwidth. The peak at $\omega\sim U$ stems from particle-hole excitations, whereas the low frequency absorption appears as a consequence of the formation of a Bose-glass at incommensurate filling.}
\label{fig:sketch}
\end{figure}

The resulting general picture of the energy absorption is depicted in Fig.~\ref{fig:sketch}.
The low frequency $0<\omega<W$, where $W \sim \textrm{max}(4J,2\Delta)$  is the effective bandwidth, is only present for incommensurate  fillings, $\nu<1$.
A second distribution  is centered at frequency $\omega= U$ and comes from particle-hole excitations
generating doubly occupied sites. The width of this absorption line is also given by $W$.

The paper is organized as follows. In Section~\ref{sec:sec2}, we present the general formalism
needed to calculate the absorption rate for  incommensurate and unit  filling. In Section~\ref{sec:sec3} and Section~\ref{sec:sec4}, we present our analytical and numerical results obtained for the random  and the quasi-periodic potential, respectively. In Section~\ref{sec:sec5}, we discuss the effects of a trapping potential. Finally in Section~\ref{sec:sec6} we provide our conclusions. A  derivation of the general formula [see Eq.~(\ref{asy_weak})] for the absorption rate valid for weak disorder is given in the Appendix~\ref{sec:sec7}).

\section{Energy Absorption: Linear Response Theory}\label{sec:sec2}

 For weak perturbations, corresponding to $\delta J \ll J$, the energy absorption rate $\dot{E}_{\omega}$ can be calculated using linear response theory.  The general formula has been first derived in Ref.\cite{iucci_shake_bosons_theory}
and in the presence of disorder it takes the form
\begin{equation}
\dot{E}_{\omega }=\frac{1}{2}\delta J_{0}^{2}\,\omega \:  \overline{\mathrm{\Im}\left[ -\chi _{K}(\omega )\right] },  \label{def}
\end{equation}
where $\chi_{K}(\omega )$ is the Fourier transform of the \emph{retarded} correlation function $\chi _{K}(t)=-i\Theta (t)\langle [ K(t),K(0)] \rangle$ of the hopping operator $K=-\sum_{j}(b_{j+1}^{\dag }b_{j}^{\pd}+\text{h.c.})$, being $\Theta(t)$ the step function.  In Eq.~(\ref{def}) the bar means average over different disorder (Sect.~\ref{sec:sec4})
or quasi-periodic (Sect.~\ref{sec:sec3}) realizations.

In general,  the calculation of the correlation function in Eq.~(\ref{def}) is a complicated many-body problem.
However, in the limit of strong repulsion where $U \gg J,\Delta$, calculations are considerably simplified by the
fact that we can accurately restrict ourselves to work within a subspace of the total Hilbert space, whose detailed
structure  depends on the filling.  In the case of an \emph{incommensurate} filling (\emph{i.e.} $\nu <1$) this subspace
can be described in terms of non-interacting fermion states (that is, Slater determinants).  For commensurate
filling (\emph{i.e.} \emph{unit} filling, $\nu =1$) we can restrict ourselves to the subspace with one particle and one
hole excitation, as described below.

\subsection*{Incommensurate filling}

For filling $\nu<1$ and large on-site repulsion $J,\Delta \ll U $, we take the hard core limit $U\rightarrow +\infty$.
Bosons are then mapped onto \emph{non interacting} spinless fermions via the Jordan-Wigner transformation:
\begin{equation}  \label{Jordan}
c_{j} = \exp\left[ i\pi\sum_{k=1}^{j-1}n_{j}\right] b_{j},
\end{equation}
where $c_j$ satisfy fermionic commutation relations $\{c_j^{\pd},c_j^{\pd}\}=0$ and $\{c_j^{\pd},c_j^{\dag}\}=1$. Under the above transformation, the Hamiltonian (\ref{Hbh}) is mapped onto the single particle Hamiltonian:
\begin{equation}  \label{h0}
H^\prime=-J\sum_j (c_{j+1}^{\dag}c_{j}^{\pd}+\text{h.c.}) + \sum_j\epsilon_{j}^{\pd} \: c_{j}^{\dag}c_{j}^{\pd},
\end{equation}
where the on-site repulsion $U$ has disappeared and the hopping operator $K$ in Eq.~(\ref{def}) is now given by
$K^\prime=-\sum_{j}(c_{j+1}^{\dag }c_{j}^{\pd}+\text{h.c.})$. Notice that the mapping of observables that are non local in space
is far less trivial, an example is the  momentum distribution studied in Ref.~\cite{egger_disorder} for disordered hard-core bosons in one dimension.

After some algebra, the absorbtion rate~(\ref{def}) becomes
\begin{equation}  \label{formula}
\dot{E}_{\omega}=\,\frac{\delta J_{0}^{2}\pi\omega}{2}\sum_{\alpha,\beta }\overline{\mathcal{K}_{\alpha\beta}\left[f(\varepsilon_{\alpha})-f(\varepsilon _{\beta})\right] \delta(\omega+\varepsilon_\alpha-\varepsilon_\beta)},
\end{equation}
where the matrix  $\mathcal{K}$ is defined as:
\begin{equation}  \label{K}
\mathcal{K}_{\alpha\beta}=|\sum_{j}\left[ \psi_{\alpha}(j+1)\psi_{\beta}(j)+\psi_{\alpha}(j)\psi_{\beta}(j+1)\right]|^{2}.
\end{equation}
In the previous expressions $\varepsilon_\alpha$ and $\psi_\alpha(j)$ are the eigenvalues and eigenfunctions of $H^\prime$, respectively. In Eq.~(\ref{formula}) $f(\varepsilon)=(\exp[(\varepsilon-\mu)/T]+1)^{-1}$ is the Fermi-Dirac distribution function at a temperature $T$ and  chemical potential $\mu$. The latter is fixed by the normalization condition $\nu=\sum_\alpha f(\varepsilon_\alpha)$.  At zero temperature the only relevant processes in Eq.~(\ref{formula}) correspond to transitions from an occupied level  (with energy $\varepsilon_\alpha < \mu(T = 0)$) to an unoccupied level ($\varepsilon_\beta > \mu(T = 0)$). In particular,  for unit filling  the absorption   (\ref{formula})  vanishes, consistently with the fact
that a Mott insulator can only absorb at much higher frequencies $\omega \sim U$.

In the absence of disorder [\emph{i.e.} for $\epsilon_i=0$ in Eq.~(\ref{h0})],  the
hopping modulation commutes with the Hamiltonian $H'$  and therefore
the absorption rate vanishes to all orders, even beyond
linear response. In the above expression, this is reflected in
that, for $\epsilon_i = 0$,  the eigenstates of $H'$ become plane waves, $\psi_k(j) \propto e^{i k j}$, with energy
dispersion $\varepsilon_k=-2J \cos k$ ($k$ being the  lattice momentum). Thus the matrix (\ref{K}) is  diagonal, \emph{i.e.} $\mathcal{K}_{kk^{\prime}}=4\delta_{k,k^{\prime}}\cos^2 k$, which, together with the factor $f(\varepsilon_{k})-f(\varepsilon_{k^{\prime}})$ in Eq.~(\ref{formula}) makes $\dot{E}_{\omega}$ vanish.

At weak disorder (\emph{i.e.} $J\gg \Delta$), the absorption rate (\ref{formula})
can be evaluated using perturbation theory (the details can be found in the Appendix), which
yields:
\begin{equation}\label{asy_weak}
\dot{E}_{\omega}=\,\frac{\delta J_{0}^{2}\pi\omega}{2M} \sum_{k,k^\prime } \frac{\overline{|V_{k-k^\prime}|^2} }{J^2}
\left[f(\varepsilon_k)-f(\varepsilon _{k^\prime})\right] \delta(\omega + \varepsilon_k -  \varepsilon _{k^\prime}  ),
\end{equation}
where $V_{k}=\frac{1}{\sqrt{M}}\sum_{j=0}^{M-1} e^{i k j}\epsilon_j$
is the  Fourier transform of the disorder potential. Equation (\ref{asy_weak}) shows that
the perturbation expansion in disorder is well defined provided the Fourier transform, $V_{k}$, is finite.

 Let us finally consider the so-called atomic limit, which corresponds to $J\ll \Delta$ (yet $\Delta \ll U$). In this
 limit,  tunneling can be neglected and the eigenstates are  given by
$\psi_m(j)=\delta_{jm}$. From Eq.~(\ref{K}) we find $\mathcal{K}_{j j^{\prime}}=1$ if $j$ and $j^{\prime}$
are nearest neighbor and zero otherwise. The absorption rate,  Eq.~(\ref{formula}), thus reduces to
\begin{equation}  \label{asy_strong}
\dot{E}_{\omega}=\,\frac{\delta J_{0}^{2}\pi\omega}{2}\sum_{r=\pm 1}\sum_{j=0}^{L-1}\overline{\left[f(\varepsilon_j)-f(\varepsilon _{j+r})\right] \delta(\omega+\varepsilon_j-\varepsilon_{j+r})}.
\end{equation}
In Sect.~\ref{sec:sec31} we shall explicitly compare the results obtained using exact numerical diagonalization with the
above results obtained both in the limit of weak (\ref{asy_weak}) and strong  (\ref{asy_strong}) disorder.
 Finally, it is important to emphasize that   Eq.~(\ref{formula}) does not account for
 particle-hole excitations which are relevant at much higher frequencies, $\omega \sim U$.
 These excitations become particularly important at unit filling ($\nu=1$),
when the strongly repulsive Bose gas becomes a Mott insulator and
the absorption at low frequency $\omega \ll U$ predicted by Eq.~(\ref{formula}) vanishes  because there are
no empty sites (\emph{i.e.} holes) in the ground state.  The contribution to the absorption
from particle-hole excitations at unit filling will  be discussed next.

\subsection*{Unit filling}

We next turn our attention to the \emph{commensurate} case with $\nu=1$. For large enough $U/J$ and $U/\Delta$,  the system becomes a bosonic Mott insulator. In Ref.~\onlinecite{iucci_shake_bosons_theory}, it was shown that for clean systems the absorption rate is zero at low frequencies and exhibits a narrow peak of width $\sim J$ centered about $\omega=U$. In this section, we consider the modifications of such peak due to a disorder or quasi-periodic potential, $\epsilon_i$.

In order to obtain the energy absorption rate within linear response, we first use
the spectral decomposition of the correlation function $\chi_K(\omega)$ in terms of the exact
eigenstates of the unperturbed Hamiltonian. This yields the following expression for the energy
absorption:
\begin{equation}  \label{Mott1}
\dot E_\omega=\delta J^2 \omega \frac{\pi}{2} \sum_{n} \overline{\left\vert\left\langle \Psi_n \right\vert K \left\vert \Psi_{0}\right\rangle\right\vert ^{2}\delta\left( \omega+E_{0}-E_{n}\right)},
\end{equation}
where  $\vert \Psi_n \rangle$ are the eigenstates of the original Hamiltonian (\ref{Hbh}) with energies $E_n$, and
$\left\vert \Psi_{0}\right\rangle =\left\vert 1, 1,\ldots,1 \right \rangle$ is the ground state in the Fock representation
corresponding to one boson per lattice site.
The low energy states  are  particle-hole excitations
$\left\vert \phi\left( m,j \right) \right\rangle = \frac{1}{\sqrt{2}}b_{m}^{\dagger}b_{m+j}\left\vert \Psi_{0}\right\rangle$
corresponding to double occupation at site $m$ and an empty site $m+j$.  These excitations are all degenerate with energy $U$ in the absence of tunneling and  (\emph{i.e.} for $\Delta=J=0$).

For finite values of $\Delta$ and $J$, the eigenstates $\Psi_n$ can be calculated using degenerate perturbation theory by writing $\left\vert \Psi_n \right\rangle=\sum_{m,j} f_{m,j} \left\vert \phi\left(
m, j \right)\right\rangle$, where the coefficients satisfy:
\begin{equation}\label{diag}
\sum_{m^\prime,j^\prime}\left\langle \phi\left(  m, j \right)  \right\vert H
\left\vert \phi\left(
m^{\prime},j^{\prime}\right)  \right\rangle f_{m^\prime,j^\prime}=E f_{m,j}.
\end{equation}
 The above matrix element and the energy absoption can be computed by
 taking into account that $\langle \phi\left(  m, j \right) \vert K \vert \Psi_{0}
\rangle= -\sqrt{2}\delta_{j,\pm 1}$, within the subspace containing just one particle
and one hole.  Hence, the matrix elements in  Eq.~(\ref{Mott1}) correspond to $\left\langle
\Psi_n \right\vert K \left\vert \Psi_{0}\right\rangle =- \sum_{m,r=\pm 1} \sqrt{2}
f_{m,r} $.

 Again, let us note that,  in the `atomic limit' where the tunneling can be neglected, Eq.~(\ref{Mott1})
simplifies considerably. Localized particle-hole excitations become the
exact eigenstates $\vert \Psi_n \rangle = \left\vert \phi\left( m,j \right) \right\rangle$ with energy $E_0+U+\varepsilon_m-\varepsilon_{m+j}$, where $E_0=\sum_{i=1}^{M} \varepsilon_i$ is the energy of the ground state for a given realization of $\epsilon_{i}$. The absorption rate (\ref{Mott1}) then takes the form:
\begin{equation}\label{atomic_Mott}
\dot{E}_{\omega}=\, \delta J_{0}^{2}\pi\omega\sum_{r=\pm 1}\sum_{m=1}^{M}\overline{ \delta(\omega-U+\varepsilon_m-\varepsilon_{m+r})},
\end{equation}
which can be readly evaluated numerically once the disorder potential is known.

\section{Results for a disorder potential}\label{sec:random}\label{sec:sec3}
In this section we assume that the on-site energy $\epsilon_i$ in Eq.~(\ref{Hbh}) are random numbers uniformly distributed within the interval $[-\Delta,\Delta]$. The absorption rate (\ref{def})  can thus be conveniently recast
as:
\begin{equation}\label{defF}
 \dot{E}_\omega=M\delta J_0^2 F,
\end{equation}
where $F$ is a dimensionless  function that can be numerically  evaluated. The results
for incommensurate and commensurate cases are described below.

\subsection*{Incommensurate filling}\label{sec:sec31}

\begin{figure}[tb]
\begin{center}
\subfigure{\includegraphics[width=3.0in]{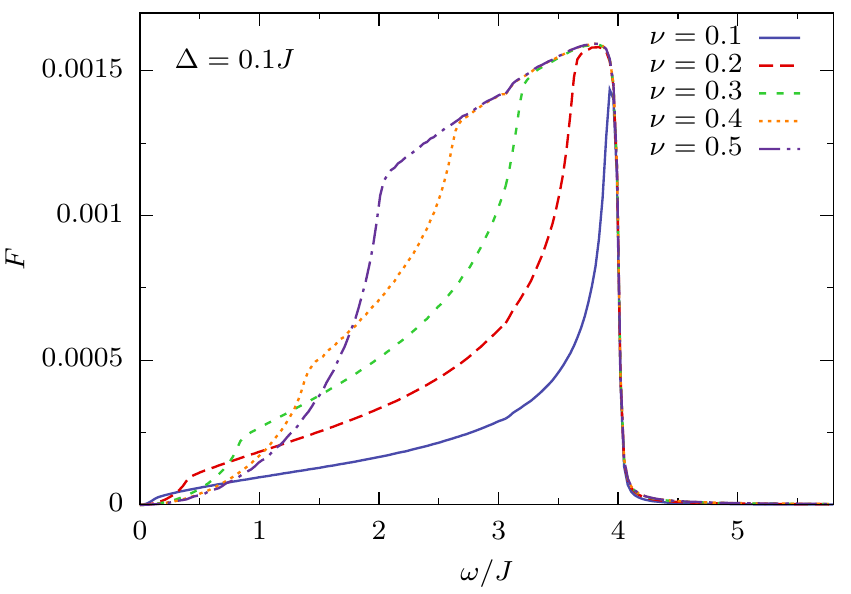}}
\subfigure{\includegraphics[width=3.0in]{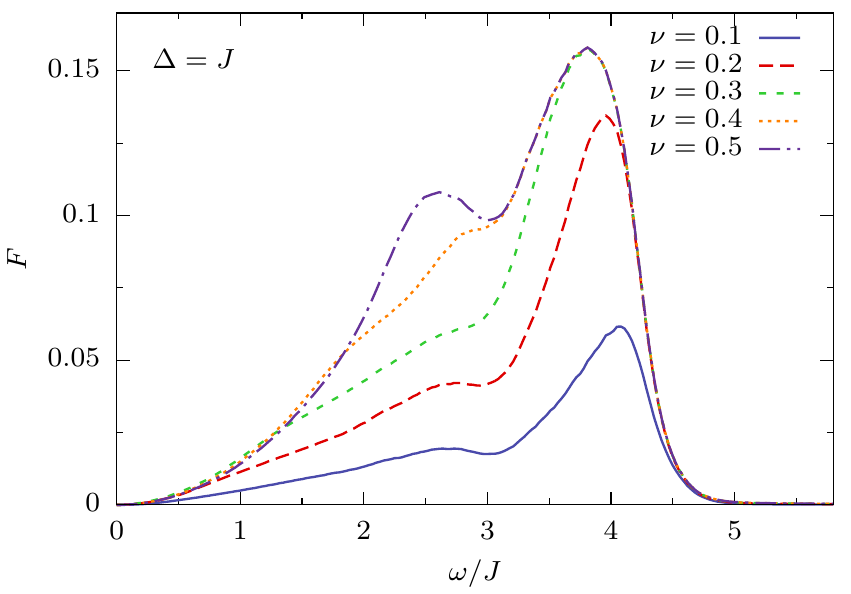}}
\subfigure{\includegraphics[width=3.0in]{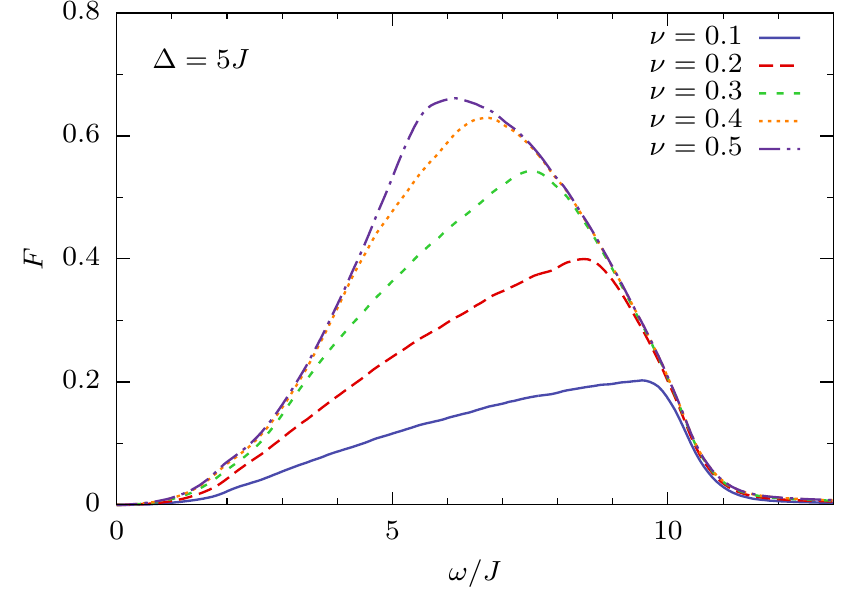}}
\end{center}
\caption{Energy absorption rate of hardcore bosons in a \emph{random} potential: the response function $F$ [see Eq~(\ref{defF})] is plotted versus modulation frequency for different filling factors and increasing values of disorder strength:  $\Delta=0.1J$ (top panel), $\Delta=J$ and $\Delta=5J$.
Calculations are done on a ring of $M=500$ lattice sites yielding negligeable finite size effects. The number of disorder realizations used was  $N_r=500$.   }
 \label{fig:fill}
\end{figure}

As stated above, the absorption rate is calculated numerically at zero temperature starting from Eqs. (\ref{formula}) and (\ref{K}). In Fig.~\ref{fig:fill} we plot the frequency dependence of the response function $F$ for \emph{different} fillings and increasing values of  disorder $\Delta=0.1 J$ (upper panel), $\Delta= J$ (central panel) and $\Delta=5 J$ (lower panel). Since, in the fermionic representation, the Hamiltonian of Eq.~(\ref{h0}) is particle-hole symmetric, the absorption rate in Eq.~(\ref{formula}) is unchanged under the transformation $\nu\rightarrow 1-\nu $, so we restrict our discussion to fillings $\nu \leqslant 1/2$.

A noticeable feature of  Fig.~\ref{fig:fill} is that the response at high frequencies is \emph{independent} of the filling factor. In this limit, the relevant processes contributing to the absorption mainly involve transitions from states far below the Fermi level (\emph{i.e.} $(\varepsilon_{\alpha} \ll \mu)$ into empty states far above it (\emph{i.e.} $(\epsilon_\beta \gg \mu)$). As a result, the Fermi-Dirac distributions in Eq.~(\ref{formula}) become irrelevant, and thus any dependence on the value chemical potential $\mu$ disappears.
\begin{figure}[tb]
\begin{center}
\includegraphics[width=\figwidth]{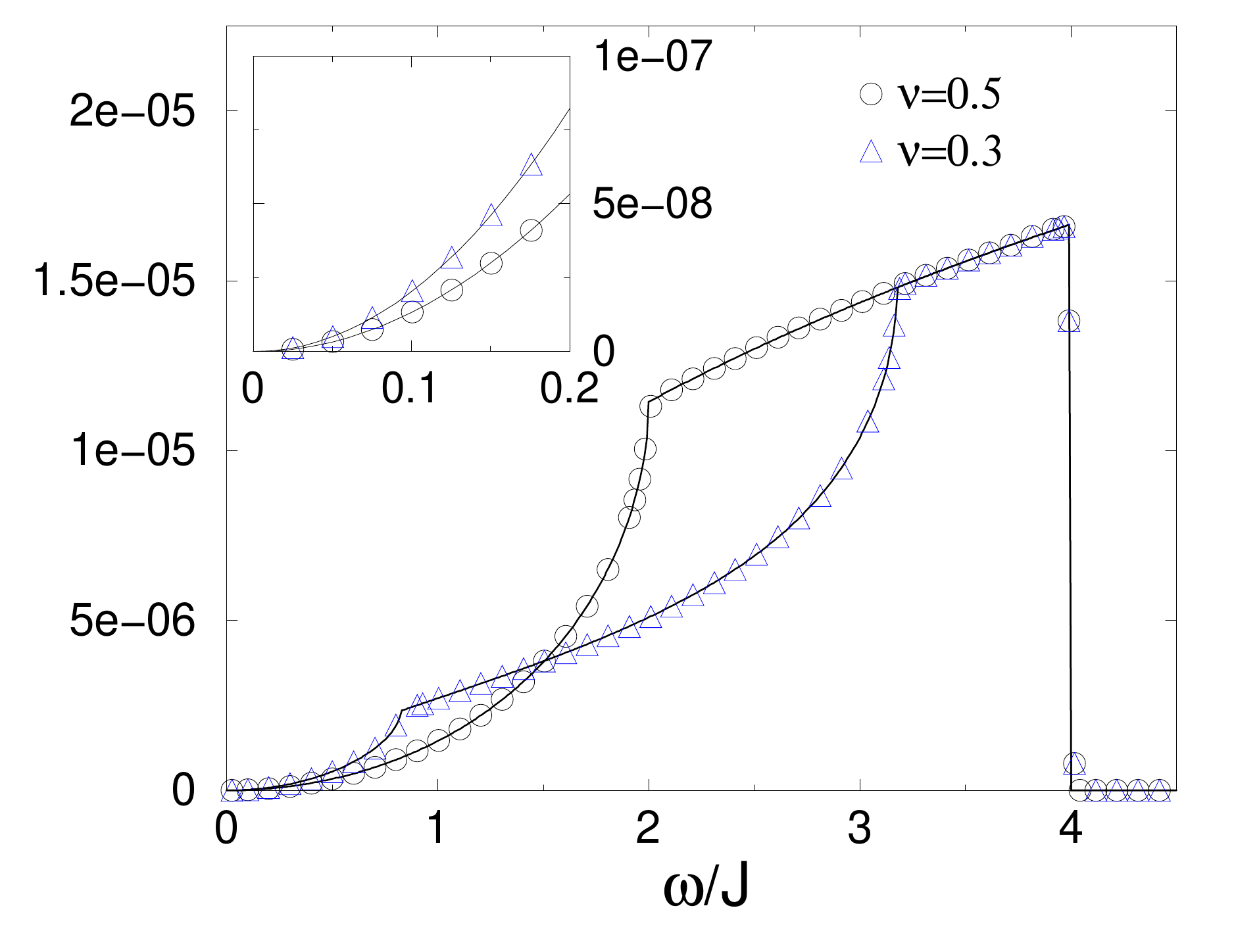}
\end{center}
\caption{
Comparison between numerics [symbols] and analytics [Eq.~(\ref{asy_weak}), solid line] for weak disorder. Here $\Delta/J=0.01$ and we consider two filling factors $\nu=0.3$ and  $\nu=0.5$. The length of the chain is $M=2000$ and the number of disorder realizations is $N_r=2000$.
The inset is a zoom of the low frequency regime, where the absorption rate is quadratic in frequency as given by Eq.~(\ref{IF})}.
\label{fig:pt}
\end{figure}

However, at low frequencies, the absorption rate  vanishes {\it quadratically} with
frequency for any strength $\Delta$ of the disorder. To understand this,   let us expand
in Eq.~(\ref{formula} the distribuntion functions, $f(\varepsilon_{\alpha})-f(\varepsilon _{\beta})\simeq (\varepsilon_{\alpha}-\varepsilon _{\beta}) \: \partial_{\varepsilon} f(\varepsilon_{\alpha})$.
Taking into account that, at zero temperature,
$\partial_{\varepsilon} f(\varepsilon) =-\delta(\mu-\varepsilon)$, we find that $F \simeq C \omega^2$, where
the constant
\begin{equation}
\label{square}
C =\frac{\pi}{2} \sum_{\alpha \neq \beta} \overline{\mathcal{K
}_{\alpha \beta} \delta(\mu-\varepsilon_\alpha)\delta(\mu-\varepsilon_\beta)}
\end{equation}
can be evaluated numerically. From Eq.~(\ref{square}) it can be seen that the constant $C$ is non zero provided  there are non-vanishing matrix elements $\mathcal{K}_{\alpha \beta}$ connecting two states $\alpha$ and $\beta$ at the Fermi level, $\varepsilon_{\alpha}=\varepsilon _{\beta}=\mu$.

In the limit of weak disorder corresponding to $\Delta \ll J$, the absorption can be evaluated using Eq.~(\ref{asy_weak}), where $\overline{|V_k|^2}=\Delta^2/3M$, as follows from the expression for the Fourier
transform  of the disorder potential, $V_{k}$.  Going to the thermodynamic limit  and introducing the density of
states $\rho(\varepsilon)=(2\pi J \sqrt{1-\varepsilon^2/4 J^2})^{-1}$, we find:
\begin{equation}  \label{weak2}
F =\frac{ \omega \Delta^2}{24 \pi J^4} \int_{a}^{b}  \frac{d\epsilon}{\sqrt{1-\epsilon^2/4J^2}\sqrt{1-(\epsilon+\omega)^2/4J^2}},
\end{equation}
where $a=\text{max}(-2J,\mu-\omega)$ and $b=\text{min}(\mu,2J-\omega)$. Here the chemical potential $\mu$ is related to the filling factor by $\mu = -2J \cos \pi \nu$.  By expanding Eq.~(\ref{weak2}) at low frequencies,
we again obtain that the quadratic behavior discussed above:
\begin{equation}  \label{IF}
F =\frac{\Delta^2 \pi}{6 J^2}\rho(\mu)^2 \: \omega^2 =\frac{\Delta^2}{24 \pi J^4}\frac{\omega^2}{\sin^2(\pi \nu)},
\end{equation}
It should be noticed that the right hand-side of Eq.~(\ref{IF}) diverges in the limit of vanishing lattice filling $\nu \rightarrow 0$  because the density of states $\rho(\varepsilon)$ has a van Hove singularity  at zero energy
in one dimension. Thus the low filling limit,  the quadratic behavior of Eq.~(\ref{IF}) is only recovered at
increasingly  low frequencies, as shown in Fig.~\ref {fig:fill}(upper panel).

In Fig.~\ref{fig:pt} we show a comparison, in the limit of weak disorder ($\Delta=0.01J$), of the numerical results (open symbols) obtained using exact diagonalization with the analytical expression of Eq.~(\ref{weak2}) (continuous lines). The agreement is indeed very good over the entire frequency range. In the inset it is demonstrated that numerical results are consistent with the quadratic behavior of Eq.~(\ref{IF}), expected at low frequencies.

\begin{figure}[tb]
\begin{center}
\includegraphics[width=\figwidth]{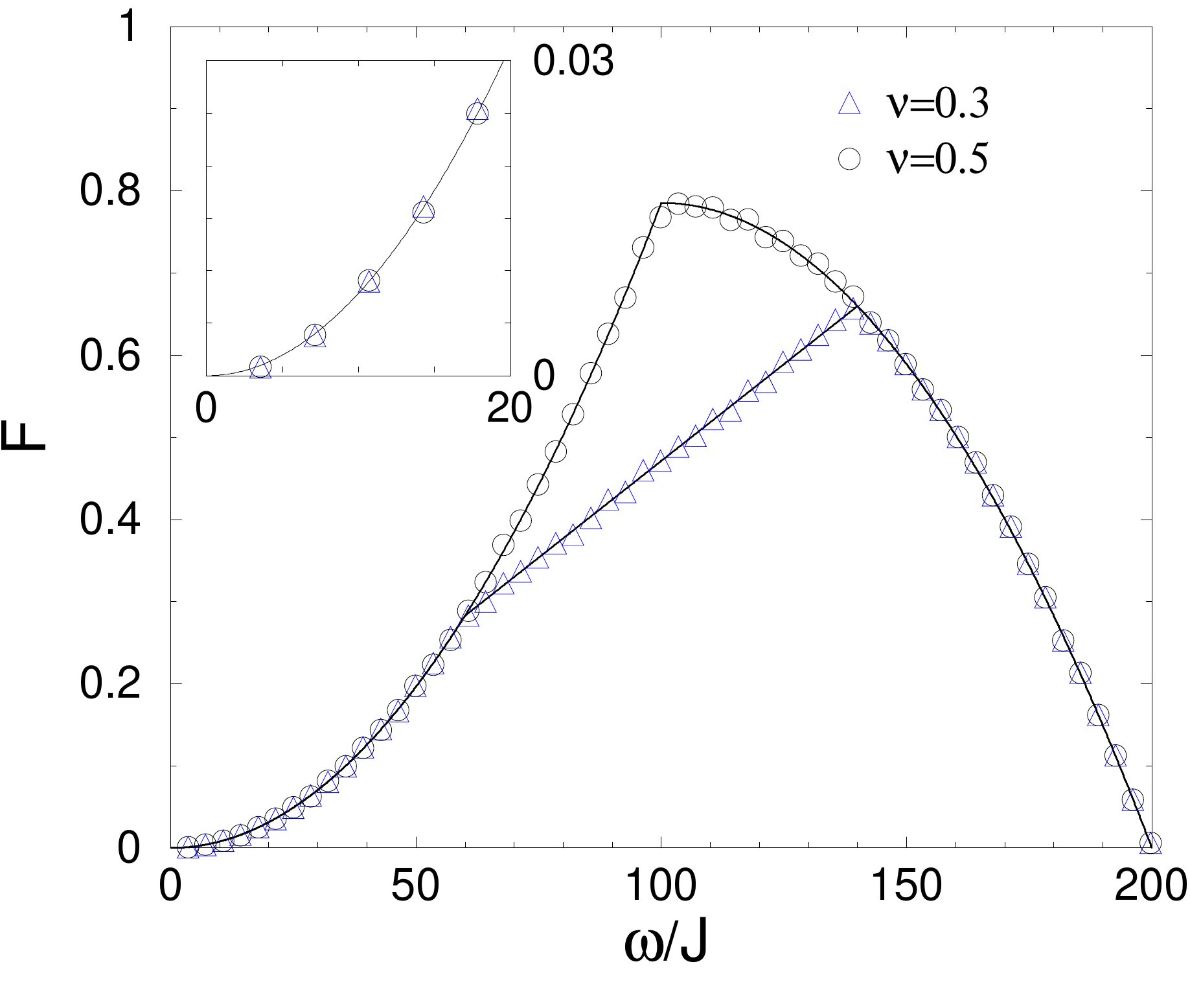}
\end{center}
\caption{
Comparison between  numerics [symbols] and analytics [solid line, Eq.\ref{stro})] for \emph{strong} disorder. The value of the disorder is $\Delta/J=100$ and the filling factors are $\nu=0.3$ and $\nu=0.5$. The length of the chain is $M=2000$ and the number of disorder realizations is $N_r=2000$. In the inset we compare our numerics with the quadratic expansion (\ref{IFstrongD}) expected at low frequency. In both cases the agreement is quite good.}
\label{fig:atomic}
\end{figure}

On the other hand, in the opposite limit of strong disorder, $\Delta \gg J$, the tunneling can be neglected. In this limit, the absorption rate can be evaluated directly from Eq.~(\ref{asy_strong}). Taking into account that the on-site energies $\epsilon_j$ at different sites are completely uncorrelated, we find
\begin{equation}\label{mia}
F=\pi \omega \int_{-\Delta}^{\overline \mu} d\epsilon \overline{\rho}(\epsilon) \int_{\overline \mu}^{\Delta} d\epsilon^\prime \overline{\rho}(\epsilon^\prime) \delta(\omega + \epsilon-\epsilon^\prime),
\end{equation}
where $\overline \mu$ and $\overline{\rho}$ are the \emph{disorder-averaged} chemical potential and  density of states, respectively. In the random potential the latter is constant and given by $\overline{\rho}(\epsilon)=1/2\Delta$, and therefore $\overline{\mu}=(2\nu-1)\Delta$. Using Eq.~(\ref{mia}), the following low-frequency behavior is
obtained:
\begin{equation}
\label{stro}
F= \frac{\pi}{4\Delta^2}\omega \left[\text{min}(\overline\mu,\Delta-\omega)-\text{max}(-\Delta,\overline\mu-\omega)\right].
\end{equation}
This behavior is exhibited by the numerics, as shown in Fig.~\ref {fig:fill} (lower panel, see  also discussion
further below). Furthermore,  in the low frequency limit Eq.~(\ref{stro}) we again recover the quadratic behavior
described above on general grounds:
\begin{equation}  \label{IFstrongD}
F=\frac{\pi}{4 \Delta^2}\omega^2.
\end{equation}
Notice however that the proportionality constant is now independent of the filling factor.
In Fig.~(\ref{fig:atomic}) a more detailed comparison of the numerics with the analytical result of Eq.~(\ref{stro}) for the limit of strong disorder is shown. In the inset, we also compare our numerical results with the quadratic behavior (\ref{IFstrongD}) expected at low frequency. In both cases the agreement is remarkably good.
%

%
\subsection*{Unit filling}

\begin{figure}[t]
\begin{center}
\includegraphics[width=\figwidth]{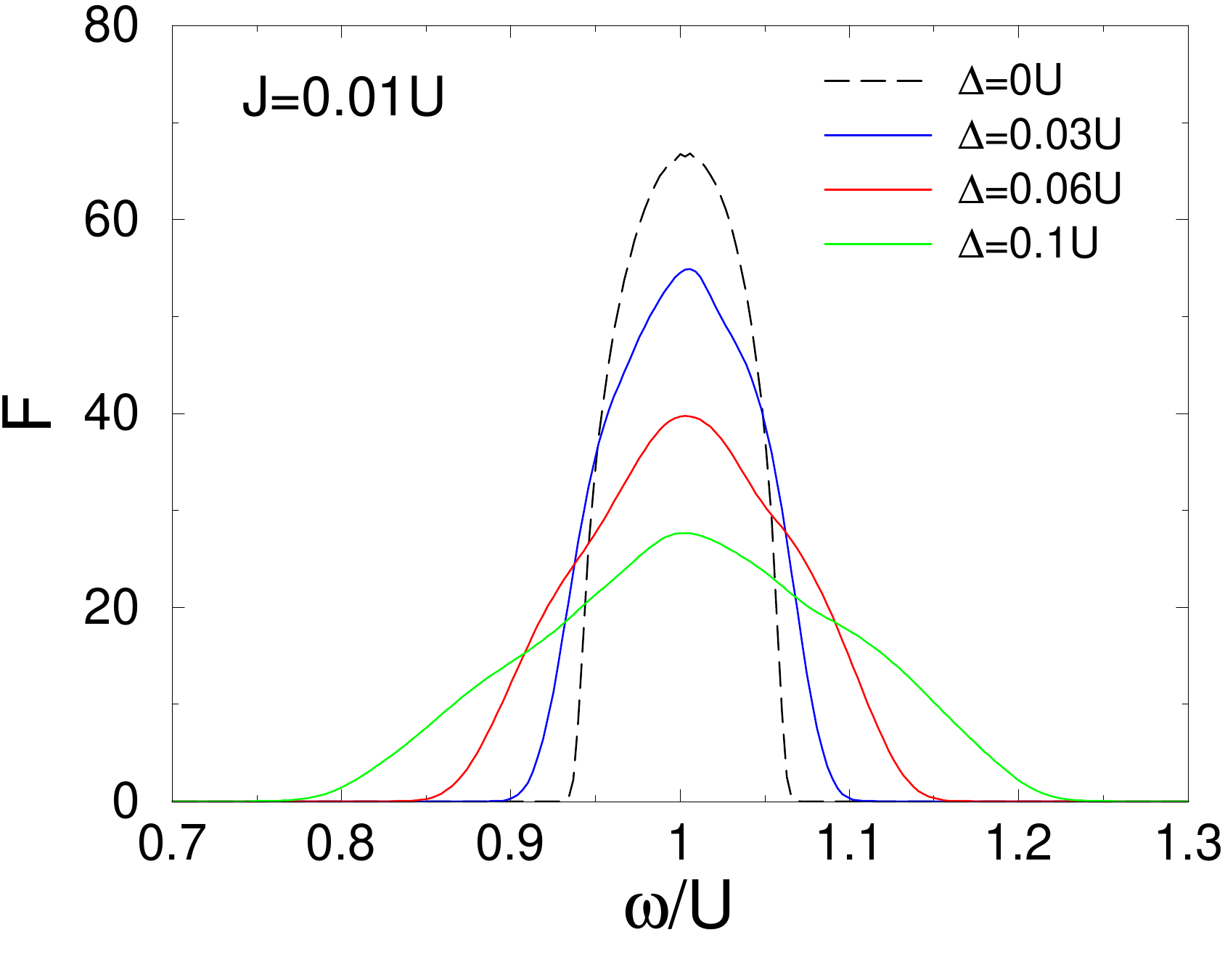}
\end{center}
\caption{Energy absorption rate in the Mott insulator phase for hardcore bosons in a random potential: the response function $F$ (see Eq.~(\ref{defF})) is plotted versus modulation frequency for fixed $J=0.01U$ and increasing values of the disorder strength $\Delta/U=0 \textrm{(dashed line)},0.03,0.06, 1$. The length
of the chain is $M=60$ and the number of disorder realizations is $N_r=200$.  A broadening of the absorption peak is observed for increasing disorder.}
\label{fig:Mott}
\end{figure}

The absorption rate for $\nu=1$ is calculated numerically starting from Eqs (\ref{Mott1}) and (\ref{diag}). The result is shown in Fig.~\ref{fig:Mott} as a function of frequency for different values of
the disorder strength $\Delta$ and $J=0.01U$.

In the absence of disorder $(\Delta=0)$ the absorption rate can be evaluated analytically
and the dimensionless function $F$ in Eq.~(\ref{defF}) is given by \cite{iucci_shake_bosons_theory}:
\begin{equation}\label{zero_dis}
F=\frac{2\omega}{3J} \left| \sin \left[ \cos^{-1}\: \left(\frac{\omega-U}{6J}\right)\right] \right|,\;\;\;\mbox {for}\, \, |\omega-U|<6J,
\end{equation}
and vanishes otherwise.  The dashed line in Fig.~\ref{fig:Mott} corresponds to our numerical result for $\Delta = 0$ which fully agrees with the above formula.

 As the strength of disorder is increased, we see
 from Fig.~\ref{fig:Mott} that the absorption spectrum becomes broader and progressively develops
a triangle-like shape as the limit $J\ll \Delta$ is approached. This behavior can be obtained analytically
starting from Eq.~(\ref{atomic_Mott}). In the $J\ll \Delta$ limit, since the on-site energies at different lattice
sites are uncorrelated,  we obtain:
\begin{equation}  \label{MottD}
F \simeq \frac{\pi \omega}{2 \Delta^2} \int_{-\Delta}^\Delta d\epsilon \int_{-\Delta}^\Delta d\epsilon^\prime \delta(\omega-U+\epsilon-\epsilon^\prime),
\end{equation}
where $\overline{\rho}(\epsilon)=1/2\Delta$ is the disorder-averaged density of states in the atomic limit
introduced above. Upon integration, Eq.~(\ref{MottD}) yields
\begin{equation}  \label{stroD}
F \simeq \frac{\pi \omega}{2\Delta^2}(2\Delta-|\omega-U|),
\end{equation}
showing that the lineshape of the absorption rate in the atomic limit is approximately triangular and vanishes at
$|\omega-U|=2\Delta$ for $J\ll \Delta$.

\section{Quasi-periodic potential}\label{sec:sec4}

In this section we assume that the external potential in  Eq. (\ref{Hbh}) is $\epsilon_j = \Delta \cos(2\pi j \sigma)$~\cite{roux_quasiperiodic_phase_diagram,deng_phase_diagram_bichromatic,roscilde_one_dimensional_superlattices}, being $\sigma$ an irrational number. This quasi-periodic potential distribution is realized experimentally by superimposing two different periodic potentials with incommensurate lattice  periods~\cite{fallani_bichromatic_shaking}. Differently from the disorder potential considered in the previous section, where all states are localized (in the thermodynamic limit) for an arbitrarily small amount of disorder, in the quasi-periodic case there is a phase transition~\cite{sokoloff_review_harper_equation} at $\Delta_c=2J$: for $\Delta<2J$ all states are \emph{extended} while for $\Delta>2J$ all states are \emph{exponentially} localized.

As described below, we  find that the absorption spectra of bosons in the quasi-periodic potential are remarkably different from the spectra described in Sect.~\ref{sec:sec3} for the bosons moving on a disorder potential.

\subsection*{Incommensurate filling}

For lattice fillings $\nu<1$  (\emph{i.e.} less than a boson per site), we calculate the absorption rate numerically
 starting from Eqs (\ref{formula}) and (\ref{K}). We restrict to zero temperature and fix $\sigma=0.77145245$ which is relevant to the experiments carried out by the Florence group.
\begin{figure}[tb]
\begin{center}
\includegraphics[width=2.8in]{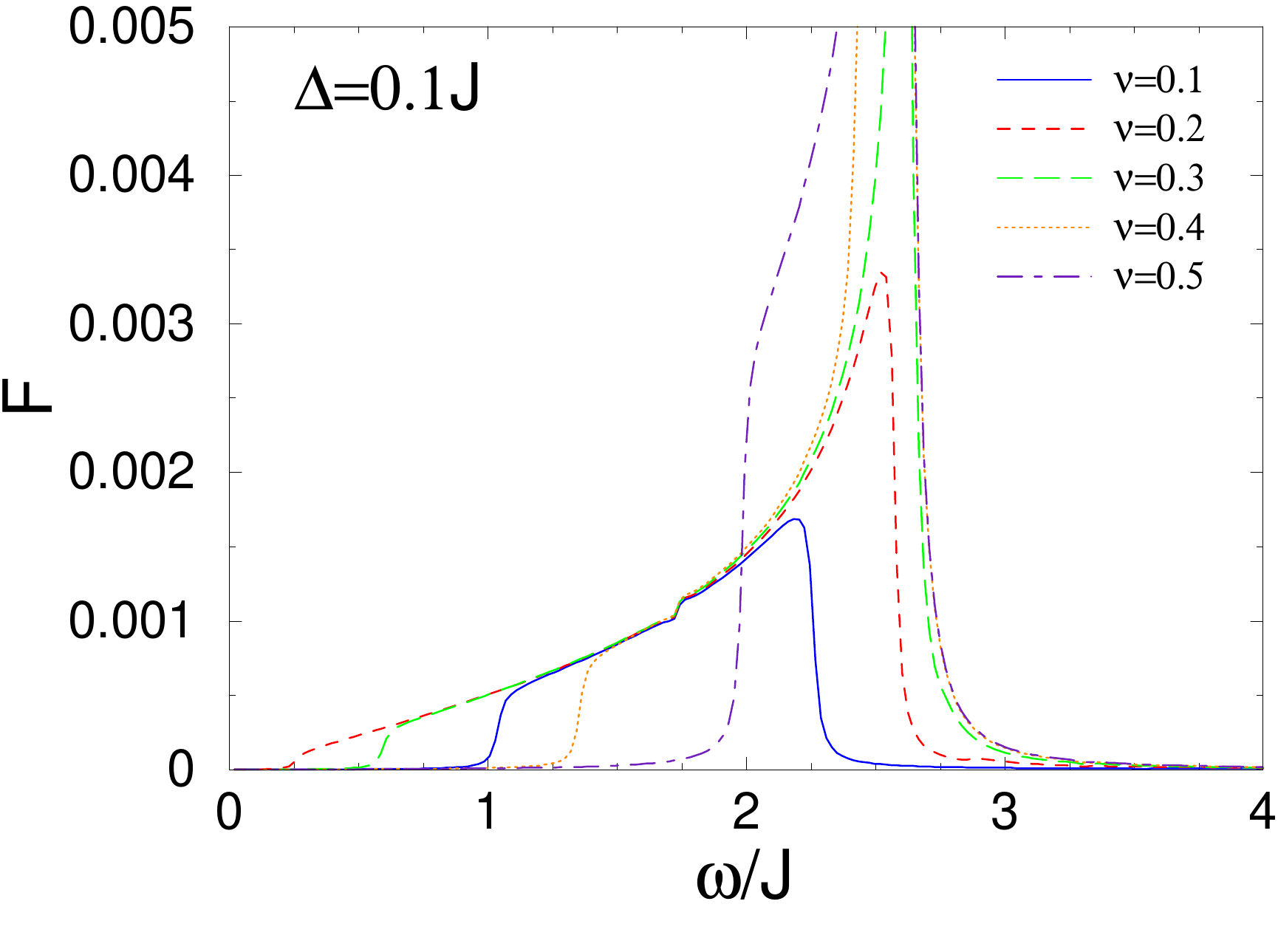}
\includegraphics[width=2.8in]{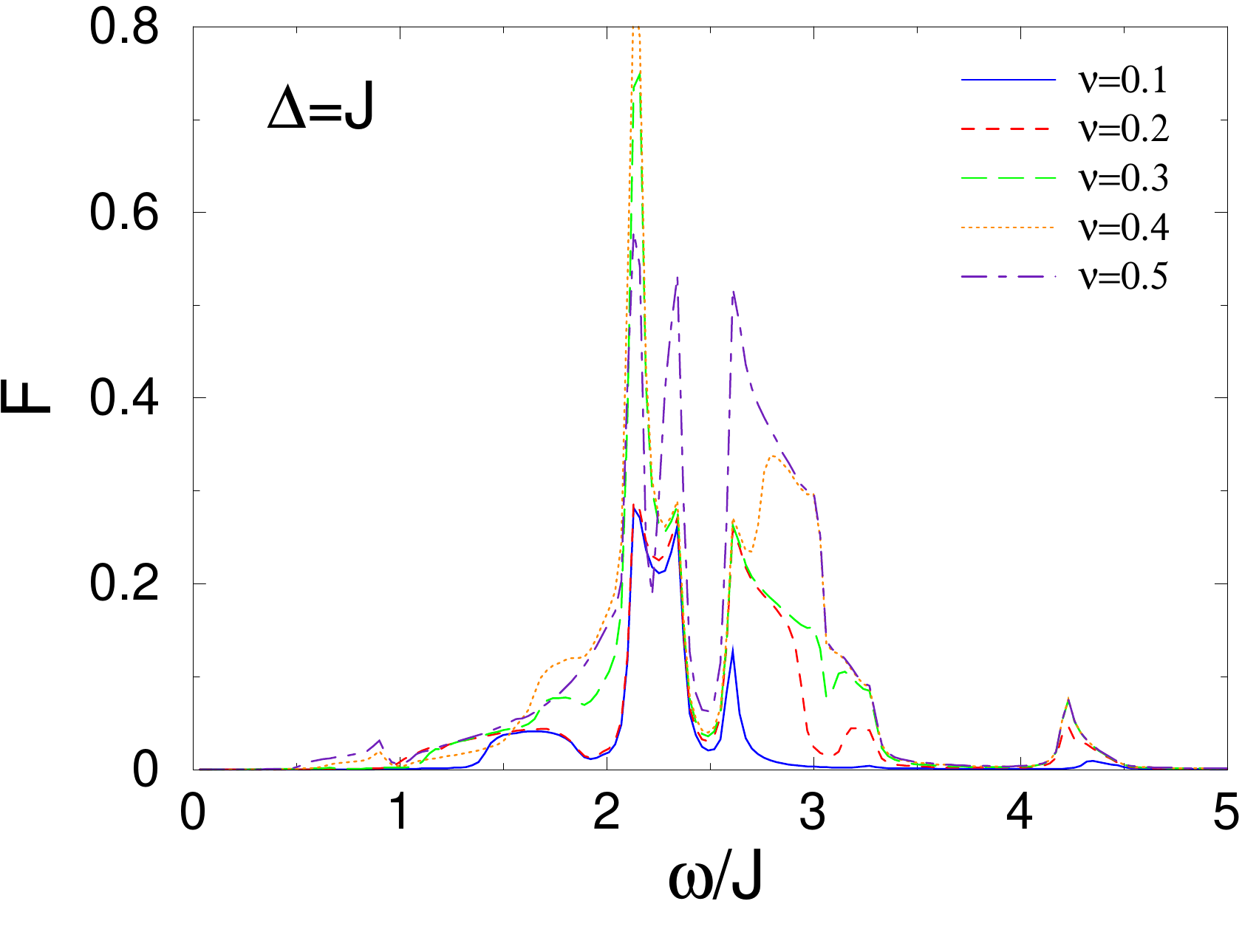}
\includegraphics[width=2.8in]{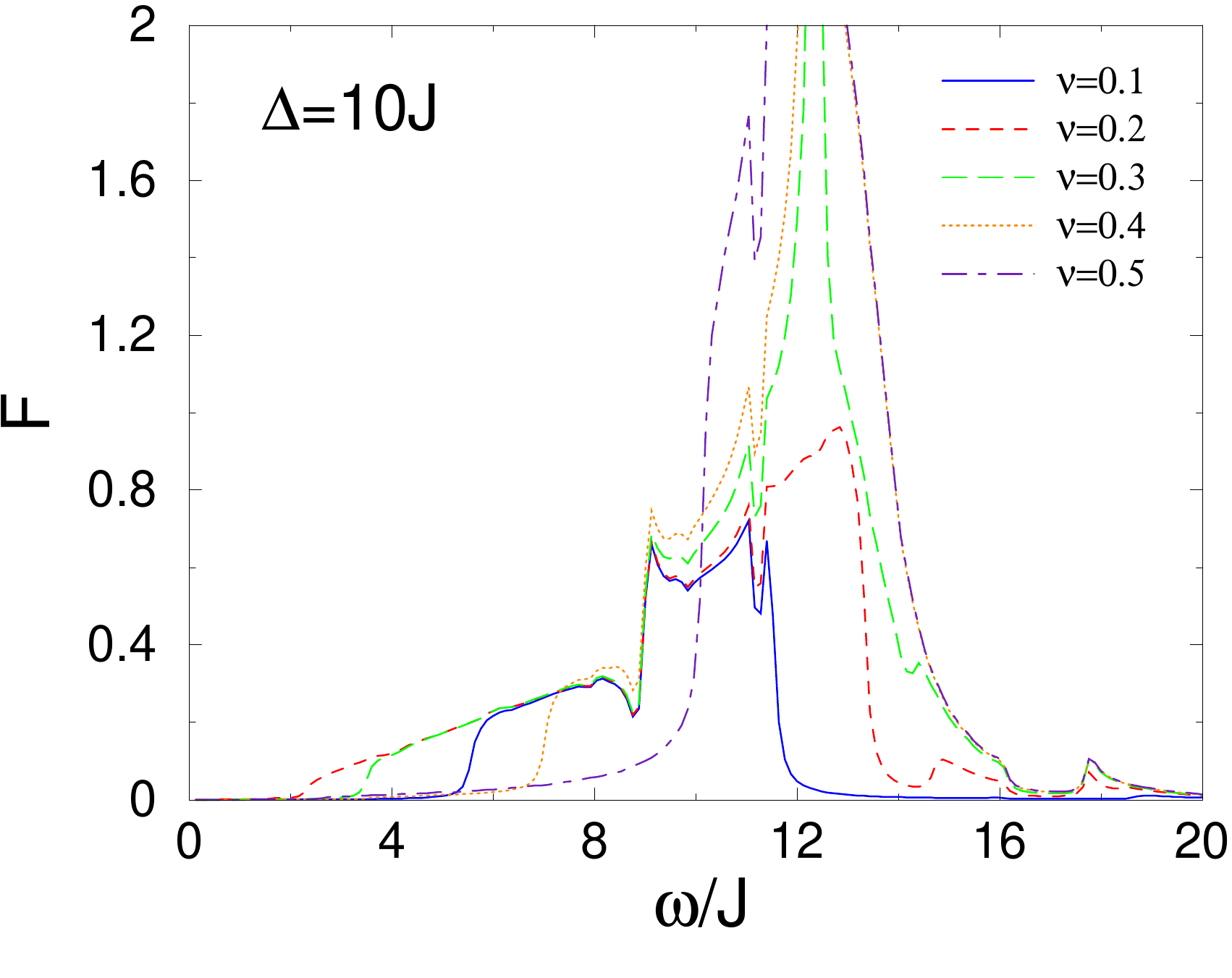}
\caption{Energy absorption rate of hardcore bosons in a \emph{quasi periodic} potential with $\sigma=0.77145245$: the response function $F$ [see Eq.~(\ref{defF})]
is plotted versus modulation frequency for different filling factors and increasing values of disorder strength:
 $\Delta=0.1J$ (top panel), $\Delta=J$ and $\Delta=10J$.
 Calculations are done on a ring of $M=1200$ lattice sites yielding almost negligeable finite size effects. The number of disorder realization is $N_r=500$. }
\label{fig:hc_quasi_disorder}
\end{center}
\end{figure}

In Fig.~\ref{fig:hc_quasi_disorder} we plot the frequency dependence of the response function (\ref{defF})
 for increasing values of the disorder strength $\Delta$. Compared to Fig.~\ref {fig:fill}, we see that the absorption spectra in quasi-periodic potential exhibit a  much richer structure.
Interestingly, the response function becomes very large (and actually diverges) when $\omega$ matches special values. Moreover, for a given filling factor, the absorption rate is non-zero only within a certain range of frequencies.

These features can again be understood in the limits of  weak and at strong disorder, where  results can be obtained analytically. In the limit  $\Delta \ll J$, the absorption rate can be calculated  starting from
Eq.~(\ref{asy_weak}) and taking the thermodynamic limit. This yields ($\varepsilon_{k} = -2J \cos k$):
\begin{equation}\label{weakTrue}
F=\frac{\pi \omega} {2} \int \frac{dk}{2\pi}\frac{dk^\prime}{2\pi} \frac{|V(k-k^\prime)|^2}{J^2}
\delta(\omega+\varepsilon_{k} +\varepsilon_{k^{\prime}}),
\end{equation}
where the integration over momenta is restricted to $[0,2\pi]$.
Using that the (modulus square of the) Fourier transform of the disorder potential
 is $|V(k)|^2=\lim_{M \rightarrow \infty} |V_M(k)|^2 $,  where
\begin{equation}\label{help}
|V_{M}(k)|^2= \frac{\Delta^2}{4M}\left | \frac{e^{i (k+k_0) M}-1}{e^{i (k+k_0)}-1} +
\frac{e^{i (k-k_0) M}-1}{e^{i (k-k_0)}-1} \right |^2,
\end{equation}
and $k_0=2 \pi \sigma$, for $M \to \infty$,  the right hand-side of Eq.~(\ref{help}) becomes
a series of delta functions:
\begin{equation}\label{diracdelta}
|V(k)|^2=\frac{\pi \Delta^2}{2}  \sum_{n=-\infty}^{+\infty}\left[\delta(k+k_0 +2\pi n)+\delta(k-k_0 +2\pi n)\right].
\end{equation}
Putting  Eq.~(\ref{diracdelta}) this result into Eq.~(\ref{weakTrue}) we obtain:
\begin{eqnarray}
F&=&\frac{\omega\pi \Delta^2}{4J^2} \int_0^{2\pi} \frac{dk}{2\pi} \delta (\omega -\varepsilon_{k}+ \varepsilon_{k+k_0}) \nonumber \\ && \Theta(\mu+\varepsilon_{k}) \Theta(\omega-\varepsilon_{k+k_0}).\label{ciao}
\end{eqnarray}
The integral in Eq.~(\ref{ciao}) can be  evaluated using the identity $a \sin k + b \cos k= c \cos(k+\gamma)$, where
$c=\sqrt{a^2+b^2}$ and $\tan \gamma=-a/b$. Since $a=\sin k_0$ and $b=1-\cos k_0$ we obtain $c=2\sin (k_0/2)$
and $\tan \gamma=-1/\tan(k_0/2)$. The zeroes of the $\delta$-function in Eq.~(\ref{ciao})
occur at $k=k_\pm=-\gamma\pm \arccos(\omega/2 J c)$. Hence, we obtain:
\begin{equation}\label{risweak}
F=\frac{\Delta^2}{4J^2}\frac{\omega/2}{\sqrt{(2 J c)^2-\omega^2}} \sum_{r=\pm} \Theta(-\xi_r) \Theta(\omega+\xi_r),
\end{equation}
where $\xi_r=- \epsilon_{k_r} - \mu = 2J \cos k_r -\mu$. Eq.~(\ref{risweak}) shows that the absorption is finite only in a range of frequencies
given by the conditions $\xi_r<0$ and $\xi_r+\omega>0$, where $\xi_r$  itself depens on $\omega$ through the
$\omega$ dependence of $k_r$. Moreover, the response diverges for $\omega=2 J c$, provided this frequency value is allowed for a given filling factor. Since  $c=2\sin \pi \sigma=1.31576$, the divergence occurs at $\omega=2.6315J$, as
found numerically and shown in Fig.~\ref{fig:hc_quasi_disorder} (upper panel).

 For  a potential strength comparable to the tuneling rate, \emph{i.e.} $\Delta \sim J$, the absorption spectrum develops very sharp peaks as shown in the central panel of Fig.~\ref{fig:hc_quasi_disorder}. These peaks gradually disappear as
$\Delta$ increases and becomes much larger than $J$. In this regime, however, the energy  absorption spectrum becomes indeed rather similar to the case of weak quasi-periodic potential, as can seen by comparing the upper and lower panels of Fig.~\ref{fig:hc_quasi_disorder}. This peculiar effect can be explained analytically starting from Eq.~(\ref{asy_strong}) which applies in the atomic limit. Introducing the variable $y=2 \pi \sigma n$, in the thermodynamic limit, we obtain:
\begin{eqnarray}
F=\omega\pi \int_0^{2\pi} \frac{dy}{2\pi} \delta (\omega + \Delta \cos y-\Delta \cos (y+2\pi\sigma)) \nonumber \\
\Theta(\mu-\Delta \cos y) \Theta(\omega+\Delta \cos y-\mu) \label{interqp}
\end{eqnarray}
The integral in Eq.~(\ref{interqp})  becomes the integral in Eq.~(\ref{ciao}) after a change of variable $k=y+\pi$. We thus obtain:
\begin{equation}\label{ris}
F=\frac{\omega/2}{\sqrt{(\Delta c)^2-\omega^2}} \sum_{r=\pm} \Theta(\mu+\Delta \cos k_r) \Theta(\omega-\Delta \cos k_r-\mu),
\end{equation}
showing that the behavior of the absorption rate at strong quasi-periodic potential can be obtained from Eq.~(\ref{risweak}) by simply replacing $2J$ with $\Delta$.

\subsection*{Unit filling}
\begin{figure}[tb]
\begin{center}
\includegraphics[width=\figwidth]{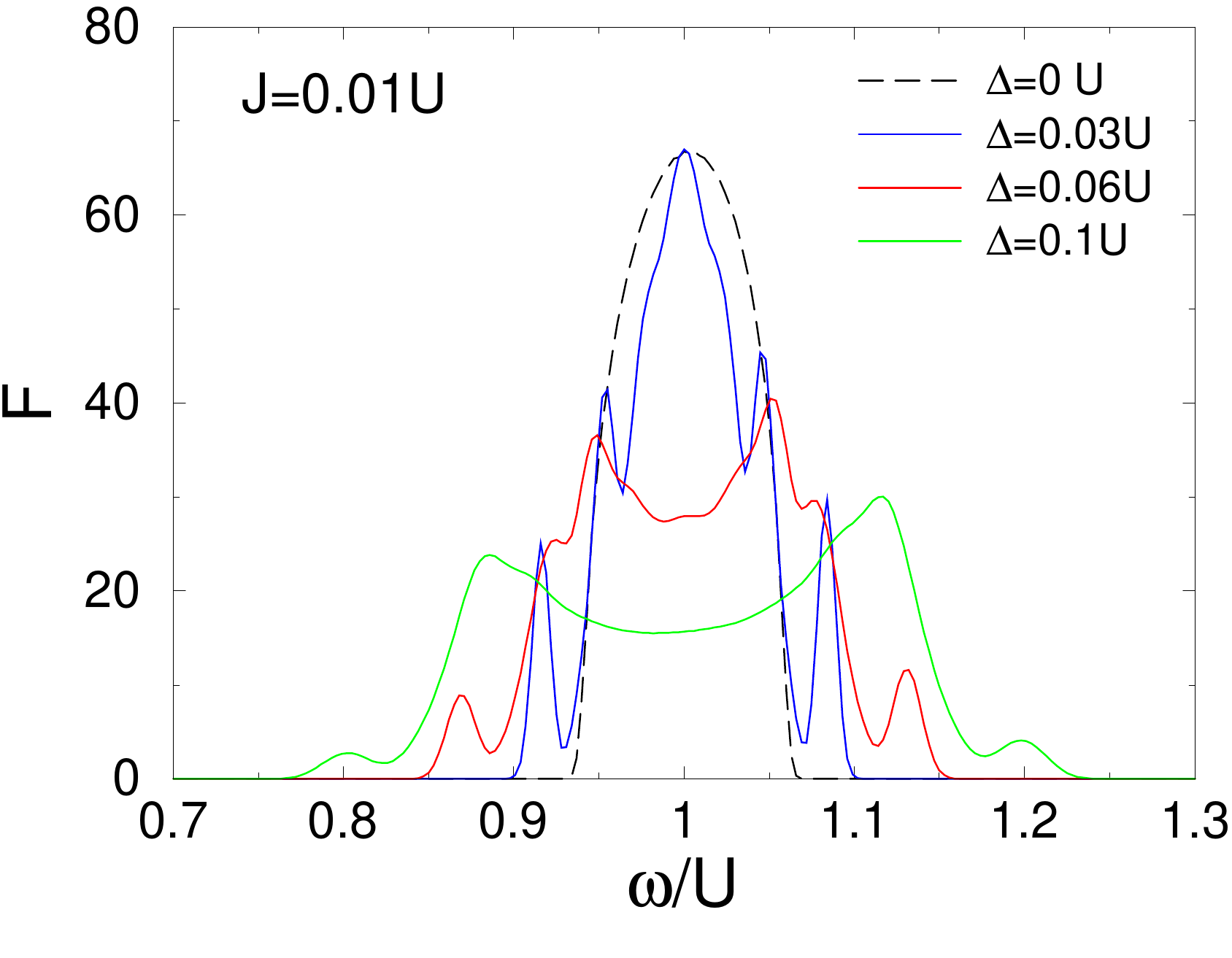}
\caption{Energy absorption in the Mott insulator phase for bosons in a \emph{quasi-periodic} potential with $\sigma=0.77145245$: the response function $F$ [see Eq.~(\ref{defF})] is plotted versus modulation frequency for fixed $J=0.01U$ and increasing disorder strength $\Delta/U=0 \textrm{(dashed line)}, 0.03, 0.06, 1$. Here convergence is achieved for system size $L=70$. As disorder increases, the response function broadens and changes convexity, as predicted by  Eq.~(\ref{asQP}).}
\label{fig:Mott_quasi_disorder}
\end{center}
\end{figure}

We have obtained the absorption spectrum at unit filling numerically using Eq.~(\ref{Mott1}) and Eq.~(\ref{diag}),
for the same value of $\sigma=0.77145245$. The result is plotted in Fig.~\ref{fig:Mott_quasi_disorder} for fixed $J/U=0.01$ and increasing values of the quasi-periodic potential strength, $\Delta$. The dashed line corresponds to the clean case, $\Delta=0$, where the absorption rate is given by Eq.~(\ref{zero_dis}).

We see that the shape of the absorption spectrum changes considerably as $\Delta$ increases.
For sufficiently weak quasi-periodic potential, $\Delta \lesssim 3J$,
the spectrum does not become broad, but instead satellite peaks appear on the sides of the central absorption
feature. For stronger quasi-periodic potentials,  the central peak disappears and the spectrum develops a two hump structure. To understand these  features, let us again focus on the `atomic limit' where the absorption rate can be obtained analitically using Eq.~(\ref{atomic_Mott}). Introducing the variable $y=2 \pi \sigma n$ and passing to the continuum limit, we obtain the result:
\begin{equation}\label{intermediate}
F=2\omega\pi \int_0^{2\pi} \frac{dy}{2\pi} \delta (\omega-U + \Delta \cos y-\Delta \cos (y+2\pi\sigma)).
\end{equation}
The integral (\ref{intermediate}) can be readily evaluated using the identity $a \sin y + b \cos y= c \cos(y+\gamma)$, where $\tan \gamma=-a/b$ and $c=\sqrt{a^2+b^2}$. From Eq.~(\ref{intermediate}) we have that $c=2 \sin \pi \sigma$ and therefore,
\begin{equation}\label{asQP}
F=\frac{2\omega}{\sqrt{(\Delta c)^2-(\omega-U)^2}},
\end{equation}
showing that the absorption rate diverges at  the the egde where $|\omega-U|=\Delta c$. This means that the shape of the absorption spectrum changes completely going from weak to strong quasi-periodic potential, as obtained numerically and shown in Fig.~\ref{fig:Mott_quasi_disorder}. Finally, upon comparing Eq.~(\ref{stroD}) and Eq.~(\ref{asQP}) we see that in a
quasi-periodic potential the absorption spectrum is  (at least in the atomic limit)  narrower because $c <2$.

\section{Effects of a parabolic trap}\label{sec:sec5}

In this section we  discuss the effects of  a harmonic trapping potential  $V (z)=m\omega_\textrm{ho}^2 z^2/2$
on the absorption spectrum. Here $m$ is the atom mass and $\omega_\textrm{ho}$ is the trapping frequency.
In this case  $V^\text{ho}_j$ in Eq.~(\ref{Hbh}) is non-zero and given by
$V_j^\textrm{ho}=\alpha^\textrm{ho}(j-M/2)^2$, where $\alpha^\textrm{ho}=m\omega_\textrm{ho}^2 d^2/2$, $d$ being the lattice period.

\begin{figure}
\begin{center}
\includegraphics[width=\figwidth]{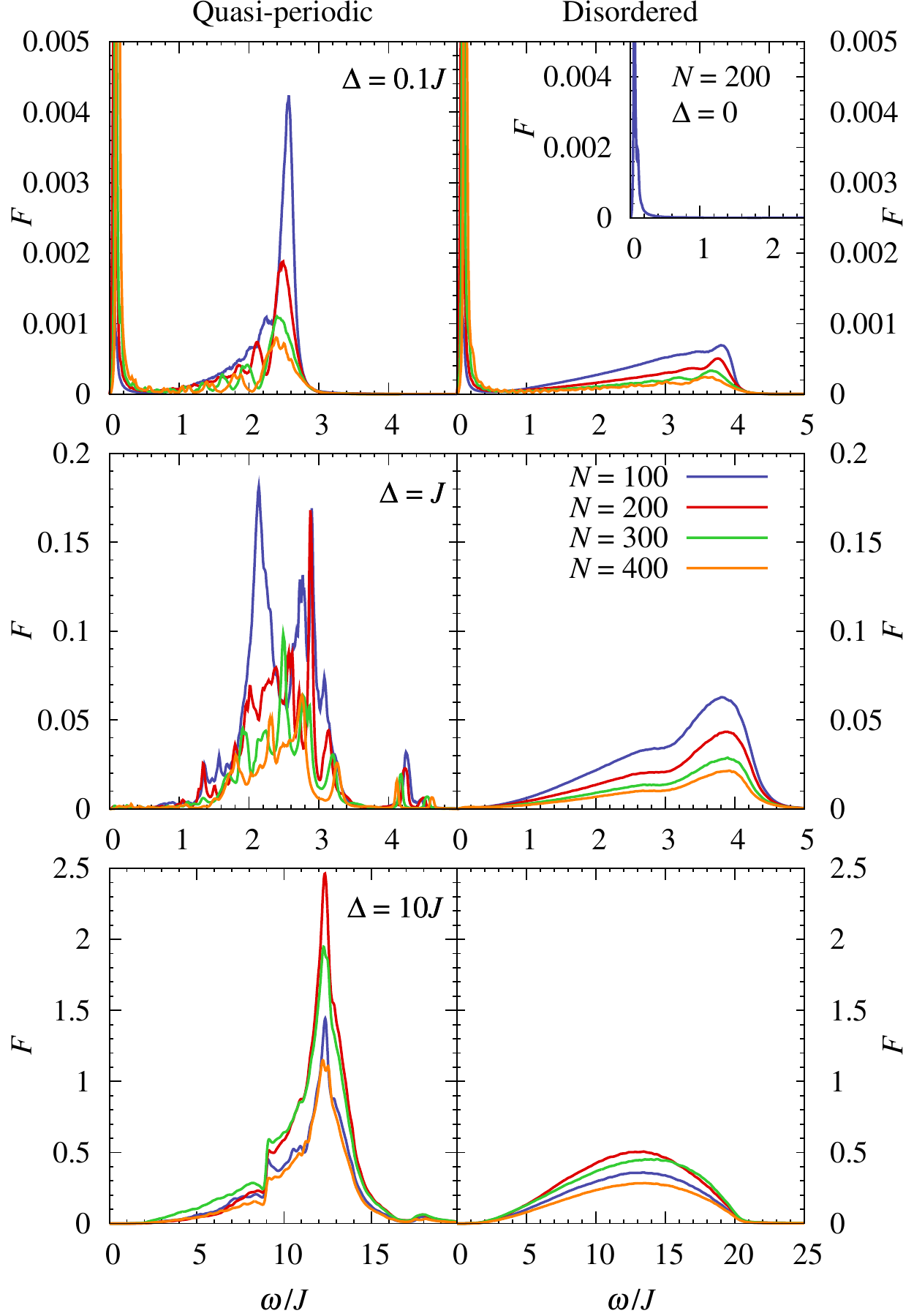}
\caption{Energy absorption rate of disordered and quasi-disordered hardcore bosons in a parabolic trap. The response function $F$ is plotted versus modulation frequency for different fillings. Upper panel: $\Delta=0.1J$, middle panel: $\Delta=J$, lower panel: $\Delta=10J$. The system size is $M=500$, the number of disorder realizations is $N_r=4000$ and $\alpha^\text{ho}=2.88\times 10^{-4}J$. The inset in the upper panel shows the low frequency absorption in the absence of disorder, which is finite for trapped gases. As $\Delta$ increases, this sharp peak fragments out and is no longer visible on the scale of the bulk contribution.  }
\label{fig:trap}
\end{center}
\end{figure}

We have repeated the calculations of the absorption spectrum including $V^\text{ho}_j$ and the result is shown in
Fig~\ref{fig:trap}. For a system of hard-core bosons in the absence of disorder or quasi-periodic potential, the trap favors the formation of a Mott insulator in the center surrounded by a superfluid region at the trap edges. This gives rise to a finite absorption
at low frequency (see inset in the upper panel),  which is related to the creation of excitations  at the edge of the trap. By contrast, in a uniform system  of hard-core bosons, as we have described in Sect.~\ref{sec:sec2} the energy absorption vanishes to all orders because the hopping operator $K$ commutes with the Hamiltonian.

Let us next consider the effect of a small amount of disorder or a weak quasi-periodic potential. Clearly the Mott insulator at the center of the trap
cannot absorb energy at low frequency, so the only contribution comes from the outer shell, where the filling factor is less than unity.
In particular the low frequency peak arising from edge excitations fragments in multiple peaks with  little spectral weight compared to the
response from the bulk discussed in Sections ~\ref{sec:random} and ~\ref{sec:sec4}.
%
%
%
%

We also see in Fig.~\ref{fig:trap} that the behavior of the absorption spectrum at frequencies close to the bandwidth crucially depends on whether the applied potential is truly random or quasi-periodic.
 Whereas the disordered case exhibits a smooth behavior in the absorption up to the bandwidth $4J$ where it falls to zero, the quasi-periodic one shows a sharp peak located at the bandwidth $4Jc$ which resembles the divergence found in the corresponding homogeneous case. Note that the position of this peak is almost independent on the number of atoms in the tube and therefore, the peak should be visible in a realistic experimental situation where an average over a multiple tube setup with variable filling is performed. The same conclusion applies to the system with strong disorder or quasi-periodic potential as can be observed in the lower panel in Fig.~\ref{fig:trap}. For $\Delta\sim J$ (Fig.~\ref{fig:trap}, middle panel) the peak structure in the quasi-periodic case is more complex, and thus the averaging procedure will produce some rounding off of the peaks. Still, the absorption can be considerably larger than in the disordered case and this difference should be clearly visible.

\section{Conclusions}\label{sec:sec6}
In conclusion we have investigated the energy absorbed by a  disordered strongly interacting  Bose gas in the presence of periodically modulated optical lattices.
For filling factor less than one, the absorption rate has been calculated exactly in the hard-core limit via the Bose-Fermi mapping. For commensurate filling, corresponding to one boson per lattice site,  the gas is a Mott insulator and can only absorbs energy at much higher frequency (of the order of the repulsive interaction $U$). The disorder induced broadening of the absorption spectrum has been calculated  by restricting to the subspace of particle-hole excitations.

We have performed extensive calculations comparing two different sources of disorder: a random potential, which is relevant for current experiments based on speckle patterns, and a quasiperiodic potential, which is obtained by superimposing two optical lattices with incommensurate periods. Our results indicate that the response of the gas to the lattice modulation significantly depends on the chosen source of randomness.

This work was partly supported by ANR grant 08-BLAN-0165-01 and by the Swiss National Fund under MaNEP and Division II. AI gratefully acknowledges financial support from CONICET and UNLP. During the initial stage of this project, GO was also supported by the Marie Curie Fellowship under contract n. EDUG-038970. MAC was supported by MEC (Spain) Grant No. FIS2007-066711-C02-02 and CSIC (Spain) through Grant No. PIE 200760/007.

\section{Appendix}\label{sec:sec7}
In this Appendix we shall derive the asymptotic formula (\ref{asy_weak}) based on perturbation theory for weak disorder.
For clarity, we rewrite the general expression (\ref{formula})
\begin{equation}\label{app1}
\dot{E}_{\omega}=\,\frac{\delta J_{0}^{2}\pi\omega}{2} \sum_{k,k^\prime } \mathcal K_{k k^\prime}
\left[f(\varepsilon_k)-f(\varepsilon _{k^\prime})\right] \delta(\omega + \varepsilon_k -  \varepsilon _{k^\prime}  )
\end{equation}
using new indices $k$ and $k^\prime$. Moreover we find convenient to  introduce the matrix elements
\begin{equation}\label{app2}
A_{k k^\prime}=\sum_{j}\left[ \psi_{k}^*(j+1)\psi_{k^\prime}(j)+\psi_{k}^*(j)\psi_{k \prime}(j+1)\right]
\end{equation}
so that $\mathcal{K}_{k,k^\prime}=|A_{k,k^\prime}|^2$.

In the absence of disorder $\Delta=0$, the eigenstates are plane waves $\psi_k^0(n)=e^{i k n}/\sqrt{L}$
with energy $\epsilon_k^0=-2J \cos k$. Therefore from Eq.~(\ref{app2}) we find
\begin{equation}\label{A0}
A^0_{k k^\prime}=2\delta_{k k^\prime} \cos k,
\end{equation}
showing that the matrix $A$ is \emph{diagonal} in momentum space. Since the matrix $\mathcal{K}_{kk^{\prime}}^0=4\delta_{k,k^{\prime}}\cos^2 k$
is also diagonal, the absorption rate (\ref{formula}) \emph{vanishes}.

For $\Delta \ll J$, we formally expand the rhs of Eq.~(\ref{app2}) in powers of the disorder strength $A_{k,k^\prime}=A^0_{k k^\prime}+A^1_{k k^\prime}+A^2_{k k^\prime}+O(\Delta^3)$, so the matrix $\mathcal K$ takes the form
\begin{align}
\mathcal K_{k,k^\prime} =&\mathcal{K}_{kk^{\prime}}^0+A^0_{k k^\prime} (A^1_{k k^\prime}+A^1_{k^\prime k}+A^2_{k k^\prime}+A^2_{k^\prime k})\label{expa}\\
& +|A^1_{k k^\prime}|^2+ \textrm{O}(\Delta^3) .\label{expa1}
\end{align}
Taking Eq.~(\ref{A0}) into account, we see that the only \emph{non-diagonal} term appearing in the expansion (\ref{expa1})
is $|A^1_{k k^\prime}|^2$, which is second order in $\Delta$.
This term can be readily evaluated
 from Eq.~(\ref{app2}) by applying first order perturbation theory for the eigenstates
\begin{equation}
\psi_k=\psi_k^0+\sum_{q \neq k} \frac{\langle  \psi_q^0|V|\psi_k^0\rangle}{\epsilon_k^0-\epsilon_q^0}\psi_q^0.
\end{equation}
After a simple algebra we obtain
\begin{equation}\label{app3}
A^1_{k k^\prime}=\frac{\langle  \psi_k^0|V|\psi^0_{k^\prime}\rangle }{\epsilon_k^0-\epsilon_{k^\prime}^0} 2(\cos k -\cos k^\prime),
\end{equation}
which is valid up to linear order in $\Delta$.
Finally, by using the dispersion relation $\varepsilon_k^0=-2J \cos k$, Eq.~(\ref{app3}) further simplifies yielding
\begin{equation}\label{ultima}
A^1_{k k^\prime}=\frac{\langle  \psi_k^0|V|\psi^0_{k^\prime}\rangle}{J}.
\end{equation}

Substituting Eq.~(\ref{ultima}) into Eq.~(\ref{app1})
and replacing the eigenstates by their zero order values $\epsilon_k=\epsilon_k^0$,
we recover the asymptotic formula (\ref{asy_weak}).


\end{document}